\documentclass[11pt,a4paper]{article}
    \usepackage{jheppub}
    
\usepackage{amsmath, amsfonts, amsthm, amssymb, graphicx}
\usepackage{color}
\usepackage[utf8]{inputenc}
\usepackage{lipsum}
\usepackage{centernot}
\usepackage{ulem}
\usepackage{feynmp}

\def\be{\begin{equation}}
\def\ee{\end{equation}}
\def\ba{\begin{eqnarray}}
\def\ea{\end{eqnarray}}
\def\beastar{\begin{eqnarray*}}
\def\eeastar{\end{eqnarray*}}       

\def\del{\partial}
\def\bdm{\begin{displaymath}}
\def\edm{\end{displaymath}}

\def\bq{\begin{quote}}
\def\eq{\end{quote}}

 at 10truept

\newcommand{\beq}{\begin{equation}}
\newcommand{\eeq}{\end{equation}}
\newcommand{\bea}{\begin{eqnarray}}
\newcommand{\eea}{\end{eqnarray}}
\newcommand{\beqa}{\begin{eqnarray}}
\newcommand{\eeqa}{\end{eqnarray}}

\newcommand{\citeseq}{\cite{KP1,KP2,KP3,KPSZ, KPS, KP4,etude}}
\newcommand{\citecc}{\cite{zeldovich,wein,pol,cliff,me}}
\newcommand{\citemon}{\cite{KS,KSq,KLS,KLuv,AKL}}

\def\ltap{\ \raise.3ex\hbox{$<$\kern-.75em\lower1ex\hbox{$\sim$}}\ }
\def\gtap{\ \raise.3ex\hbox{$>$\kern-.75em\lower1ex\hbox{$\sim$}}\ }
\def\gl{\ \raise.5ex\hbox{$>$}\kern-.8em\lower.5ex\hbox{$<$}\ }
\def\roughly#1{\raise.3ex\hbox{$#1$\kern-.75em\lower1ex\hbox{$\sim$}}}

\title{Monodromy inflation  and an emergent mechanism for stabilising the cosmological constant}

\author{Antonio Padilla} 
\emailAdd{antonio.padilla@nottingham.ac.uk}
\affiliation{School of Physics and Astronomy, 
University of Nottingham, Nottingham NG7 2RD, UK}

\date{\today}

\abstract{
We show that a pair of field theory monodromies in which the shift symmetry is broken by small, well motivated  deformations, naturally incorporates a mechanism for cancelling off radiative corrections to the cosmological constant.  The lighter monodromy sector plays the role of inflation as well as providing a rigid degree of freedom that acts as a dynamical counterterm for the cosmological constant. The heavier monodromy sector includes a rigid dilaton that forces a global constraint on the system and the cancellation of vacuum energy loops occurs at low energies via the sequestering mechanism. This suggests that  monodromy constructions in string theory could be adapted to incorporate mechanisms to stabilise the cosmological constant in their low energy descriptions.
}

\begin{document}
\maketitle

\section{Introduction}
Perhaps the simplest, most calculable, models of early universe inflation are those with superplanckian field excursions \cite{chaotic,Freese}. Typically these  give rise to  large primordial tensor fluctuations that could be detected by forthcoming polarization maps of the cosmic microwave background (CMB). However, pushing the inflaton field to such large values is a challenge for model builders keen to protect slow roll from ultra-violet corrections to the theory. Within string theory, monodromy inflation offers a promising solution to this problem \cite{mon1,mon2}. A field theory version of this has been developed in a series of recent papers \citemon~(see also \cite{Marchesano,Hebecker}), whereby a four-form field strength has a bilinear mixing with a pseudo-scalar.  Control of the effective inflaton potential stems from a $U(1)$ gauge symmetry in the four-form sector, as well as   a (discrete) shift symmetry for the axion.

A seemingly unrelated question is that of the cosmological constant, or equivalently, vacuum energy, which is radiatively unstable \citecc. Indeed, applying standard quantum field theory methods,  radiative corrections to vacuum energy scale like the cut-off of the effective field theory (EFT) to the fourth power rendering it extremely sensitive to ultra-violet physics. This is problematic because the scale of the observed cosmological constant  lies at least sixty orders of magnitude below the scale of  current collider experiments. Within a standard semiclassical framework in which quantum matter is minimally coupled to classical General Relativity, this represents  a startling failure of the so-called naturalness paradigm \cite{Giudice}. One mechanism for alleviating this problem and restoring naturalness has been dubbed {\it vacuum energy sequestering} \citeseq.  The mechanism includes new rigid degrees of freedom that force a cancellation, or better, a {\it decapitation} \cite{decap,selftun},  of radiative corrections to the vacuum energy. Consistent with the notion that relevant operators cannot be predicted in effective field theory, existing models of sequestering make no prediction for the renormalised value of the cosmological constant, but they do render it radiatively stable.  This places it on the same footing as, say,  the electron mass whose mass is protected from radiative corrections by chiral symmetry in the massless limit \cite{thooft}.

The purpose of this paper is to demonstrate that field theory models of monodromy such as \citemon  ~naturally incorporate the sequestering mechanism at low energies. To achieve this we need at least two independent monodromies, operating at hierarchically different scales and  mixing only through gravity. The lightest of these will also give rise to monodromy inflation and the heavier  to a rigid dilaton whose local fluctuations are suppressed on the scales of interest. All of this suggests that string theory, with its capacity for generating monodromy, could also have a built in mechanism for stabilising the cosmological constant.

\section{Field theory monodromy}
We begin with the prototypical model of flux-monodromy inflation introduced by Kaloper and Sorbo \citep{KS,KSq,KLS}, building upon earlier work of Dvali \cite{gia1,gia2},
\be \label{KS}
{\cal L}=-\frac{1}{2. 4!} F_{\mu\nu\alpha\beta}^2-\frac12 (\partial \phi)^2+\frac{m}{4!} \phi \frac{\epsilon^{\mu\nu\alpha\beta}}{\sqrt{-g}} F_{\mu\nu\alpha\beta}
\ee 
where $F_{\mu\nu\alpha\beta}=4 \partial_{[\mu} A_{\nu\alpha\beta]}$ is a four-form field strength. We can readily rewrite this theory in terms of pseudo-scalars only, integrating out the four-form field strength and replacing it with its magnetic dual. The result is\footnote{To pass from \eqref{KS} to \eqref{action:phi-q} we add a continuous  Lagrange multiplier field, $Q$,  to \eqref{KS}  via a term $Q \frac{\epsilon^{\mu\nu\alpha\beta}}{\sqrt{-g}}  \left(  F_{\mu\nu\alpha\beta}- 4 \partial_{[\mu} A_{\nu\alpha\beta]}\right)$, along with additional functional variation over $F_{\mu\nu\alpha\beta}$ and $Q$. Since $F_{\mu\nu\alpha\beta}$ now enters algebraicly and is at most quadratic, we can integrate it out exactly, resulting in \eqref{action:phi-q}. }
\be \label{action:phi-q}
{\cal L}=-\frac12 (\partial \phi)^2-\frac{m^2}{2} \left(\phi+\frac{Q}{m} \right)^2+\frac{1}{3!} \frac{\epsilon^{\mu\nu\alpha\beta}}{\sqrt{-g}} A_{\mu\nu\alpha} \partial_\beta Q
\ee 
where the expection value of the Lagrange multiplier, $Q$, is quantised in units of the membrane charge $q$, $\langle Q\rangle=2\pi Nq$ \cite{PS,BP}.  The action \eqref{action:phi-q} is manifestly invariant under a discrete gauge symmetry
\be \label{symmetry}
\phi \to \phi+2\pi f, \qquad Q \to Q-2\pi q
\ee
where $f=q/m$ measures the periodicity of the pseudoscalar. This gauge symmetry protects the low energy theory from large corrections, both perturbative and non-perturbative, allowing us to reliably  realise chaotic inflation at super-Planckian field values.

We can ease the tension with observational bounds on the tensor-scalar ratio for primordial fluctuations by exploiting EFT corrections to this theory. As shown in \citep{AKL}, by careful application of naive dimensional analysis (NDA) \citep{Georgi,Manohar},  the appropriate factors of $4\pi$ allow one to probe the higher derivative operators in the EFT expansion without going beyond the cut-off.  In particular, taking care to include the correct symmetry factors \citep{AKL}, these corrections take on the following generic form
\be
c_{n_1 n_2} \mu^4  \frac{\left[ \frac{(\partial \phi)^2}{2\mu^4} \right]^{n_1}}{ n_1 !} \frac{\left[ \frac{m \phi+Q}{\mu^2} \right]^{n_2}}{n_2 !}, \qquad \mu=\frac{M}{\sqrt{4 \pi}}
\ee
where $M$ is the cut-off, and $c_{n_1 n_2} \sim {\cal O}(1)$. The non-negative integers $n_1, n_2$ satisfy $2n_1+n_2 \geq 3$ with the gauge symmetry \eqref{symmetry} dictating the precise form of these interactions.

Strongly coupled dynamics now yields a theory of k-inflation \cite{kinf} with flattening of the effective potential. We refer the reader to \cite{AKL} for further details and predictions for the tensor-scalar ratio and non-Gaussianity  that can be compatible with CMB bounds, yet on the brink of being observed.  We emphasize that the EFT description remains valid  at strong coupling in a window $\mu^2 < m\phi+Q<M^2$, thanks to the extra powers of $4 \pi$. This is in contrast to many other applications of higher derivative operators  in cosmological models (see \cite{ippo} for a critique).
\section{From monodromy to sequestering}
In monodromy inflation, reheating can occur  by coupling the axion to a gauge sector in the usual way, $\frac{\phi}{f}{\textrm Tr} G \wedge G$. Non-perturbative corrections now generate a periodic potential for $\phi$. This scenario has been exploited in \cite{irat} to develop a sequestering set-up with a landscape of radiatively stable vacua.  Here we explore a different scenario, in which reheating occurs via interactions that break the discrete shift symmetry \eqref{symmetry}. For example, consider a coupling  $g \phi^2 h^2$ to  some massive scalar $h$ that itself couples to the Standard Model (in principle $h$ could even be the Higgs).  $g$ is a technically natural parameter. Coupling axions to an external Higgs-like sector,  breaking the shift symmetry along the way, is reminiscent of so-called relaxion models \cite{relaxion}. There the technically natural coupling is taken to be extremely small which is problematic for stringy realisations (see e.g. \cite{runaway}).  That will not be the case for us. What is important for us is that loops of $h$ now generate a  symmetry breaking potential, and in particular  a mass term that goes as $ \bar m^2 \phi^2$, where $\bar m \sim g m_h$ and $m_h$ is the scalar mass.  Provided $\bar m \ll m$, we do not expect a significant deformation of the inflationary dynamics. It follows that as long as the scalar mass lies below the scale of inflation (as it must anyway for efficient reheating)  we can happily tolerate any $g \lesssim {\cal O}(1)$.

To explore what happens when the gauge symmetry \eqref{symmetry} is broken explicitly in this way, let us simply deform the original Kaloper Sorbo model by the mass term described above, specifically, 
\be
{\cal L}=-\frac{1}{2. 4!} F_{\mu\nu\alpha\beta}^2-\frac12 (\partial \phi)^2+\frac{m}{4!} \phi \frac{ \epsilon^{\mu\nu\alpha\beta}}{\sqrt{-g} } F_{\mu\nu\alpha\beta}-\frac12 \bar m^2 \phi^2
\ee 
where $\bar m \ll m$ now encodes the small symmetry-breaking parameter.  This will be sufficient for elucidating the emergent mechanism for stabilising vacuum energy, so we will not include any explicit couplings between the inflaton and Standard Model fields in our  subsequent analysis. To determine the structure of the EFT corrections to this deformed theory, whilst retaining control of our power counting, it is convenient to think of $\bar m$ as a spurion, transforming under the gauge symmetry \eqref{symmetry} as  $\delta \bar m=-2\pi f \bar m /\phi$. Applying NDA as before, only now including the spurion, we find that the EFT corrections take the generic form\footnote{To apply NDA , we trade $\phi \to  \frac{4\pi \phi}{M}, \partial \to \frac{\partial}{M}, m\to \frac{m}{M}, Q \to \frac{4\pi Q}{M^2}$ and multiply the whole interaction by a factor of $\frac{M^4}{(4\pi)^2}$ \citep{Georgi,Manohar,AKL}. For the spurion we apply the rule $\bar m \to \frac{\bar m}{M}$, as with the other mass parameter. }
\be \label{dEFTcorrections}
c_{n_1 n_2 n_3} \mu^4 \frac{\left[ \frac{(\partial \phi)^2}{2\mu^4} \right]^{n_1}}{ n_1 !} \frac{\left[ \frac{m \phi+Q}{\mu^2} \right]^{n_2}}{n_2 !} \frac{\left[ \frac{\phi}{\nu } \right]^{n_3}}{n_3 !}, \quad \mu=\frac{M}{\sqrt{4 \pi}}, ~\nu=\frac{\mu^2}{\bar m}
\ee
where $c_{n_1 n_2 n_3} \sim {\cal O}(1)$ and the non-negative integers $n_1, n_2, n_3$ satisfy $2n_1+n_2+n_3 \geq 3$.  As long as $\bar m \ll m$, the inflationary dynamics is essentially the same as in \citep{AKL}, with small corrections.

Let us now imagine that we have a second monodromy sector, with a four-form $\hat F$ and a pseudo scalar $\hat \phi$, only this time we deform it gravitationally.
\be
\hat {\cal L}=-\frac{1}{2. 4!} \hat F_{\mu\nu\alpha\beta}^2-\frac12 (\partial \hat \phi)^2+\frac{\hat m}{4!} \hat \phi \frac{\epsilon^{\mu\nu\alpha\beta} }{\sqrt{-g}}\hat F_{\mu\nu\alpha\beta}+\frac12 \hat g^2 R \hat \phi^2
\ee 
where $R$ is the Ricci scalar and $\hat g \lesssim 1$. Such a deformation introduces a dynamical dilaton, prevalent in string theory and is consistent with the notion that quantum gravity should ultimately break any of the remaining global shift symmetries. Integrating out the four form so that we trade it for its magnetic dual, $\hat Q$, we find
\be 
\hat {\cal L}=-\frac12 (\partial \hat \phi)^2-\frac{\hat m^2}{2} \left(\hat \phi+\frac{\hat Q}{\hat m} \right)^2+\frac{1}{3!} \frac{\epsilon^{\mu\nu\alpha\beta}}{\sqrt{-g}} \hat A_{\mu\nu\alpha} \partial_\beta \hat Q+\frac12 \hat g^2 R \hat \phi^2
\ee 
Now if we assume that $\hat m$ lies  above the cut-off, $M$, we can decouple the local fluctuations in $\hat \phi$~\footnote{We might worry that the curvature term renormalises the effective mass for $\hat \phi$. However, as long as we insist on restricting attention to curvatures lying at or below the cut-off scale, this is not an issue.}. This forces $\hat \phi $ to lie at the minimum of its effective potential, or in other words $\hat \phi =-\hat Q/\hat m +{\cal O}(R/\hat m^2) \approx -\hat Q/\hat m$.

Bringing everything together,  including the Einstein Hilbert term and a Lagrangian for the Standard Model matter fields, we arrive at the following low energy effective theory valid below some cut-off scale, $M$,
\begin{multline} \label{seqaction6}
S=\int d^4 x \sqrt{ -g} \Bigg[ -\frac12 (\partial \varphi)^2-\frac{m^2}{2} \varphi^2+\frac{1}{3!} \frac{ \epsilon^{\mu\nu\alpha\beta}}{\sqrt{-g}} A_{\mu\nu\alpha} \partial_\beta Q
  - \mu^4  {\cal F} \left(\frac{ \varphi}{ \nu} -\frac{Q}{m \nu}, \frac{m \varphi}{\mu^2}, \frac{(\partial \varphi)^2}{2 \mu^4} \right) \\
+\frac{M_g^2}{2} \left(1+  \left(\frac{\hat Q}{\hat m \hat \nu} \right)^2 \right) R   + {\cal L}_m+\frac{1}{3!} \frac{\epsilon^{\mu\nu\alpha\beta} }{\sqrt{-g}}\hat A_{\mu\nu\alpha} \partial_\beta \hat Q \Bigg] 
\end{multline}
Here we have rewritten our inflationary monodromy sector in terms of the gauge invariant scalar $\varphi=\phi+\frac{Q}{m}$, with ${\cal F}$ containing the leading order symmetry breaking deformation and all the EFT corrections.  Indeed, the first line of this action correspond to the model of flux-monodromy inflation proposed in \cite{KS,KSq, KLS}, with a gauge symmetry-breaking deformation and EFT corrections of the form of \eqref{dEFTcorrections}.  We identify the strong coupling scale $\mu=M/\sqrt{4\pi}$ lying below the cut-off. Strongly coupled inflationary dynamics along the lines proposed in \citep{AKL} occurs when $\mu^2 < m\phi+Q<M^2$. To ensure that the inflationary behaviour is not destabilised by the symmetry breaking parameters we further assume that $\nu \gg \mu^2/m$. The last line of \eqref{seqaction6} includes the Einstein-Hilbert action along with  dilaton couplings, with the heavy dilaton held rigid below the cut-off.  The gravitational coupling  is assumed to be $M_g \sim M_{Pl}$, and we have introduced the ultra-violet scale $\hat \nu =M_g/g \gtrsim M_g$. ${\cal L}_m={\cal L}_m(g^{\mu\nu}, \Phi)$ corresponds to the Lagrangian for Standard Model matter fields minimally coupled to the metric $g_{\mu\nu}$.

We shall now demonstrate that this effective theory contains a mechanism for stabilising radiative  corrections to vacuum energy. As we will see, that mechanism is essentially sequestering \citeseq. To proceed, we compute the corresponding field equations
\begin{eqnarray}
&& \del_\mu Q =\del_\mu \hat Q=0 \label{Aeqns} \\
&&\frac{1}{4!}\epsilon^{\mu\nu\alpha\beta} F_{\mu\nu\alpha\beta}=\sqrt{-g} \frac{\mu^4}{m\nu} {\cal F}_1 \label{Qeqn}\\
&&
 \frac{1}{4!}\epsilon^{\mu\nu\alpha\beta} \hat  F_{\mu\nu\alpha\beta}=\sqrt{-g}R\frac{M_g^2 \hat Q}{(\hat m \hat \nu)^2}  \label{Qeqns}\\
&& \nabla_\mu \left ((1+{\cal F}_3) \nabla^\mu \varphi\right)-m^2 \varphi-\frac{\mu^4}{\nu} {\cal F}_1-m\mu^2{\cal F}_2 =0 \qquad \label{vphieqn} 
\end{eqnarray}
and
\begin{multline}
M_g^2  \left[1+  \left(\frac{\hat Q}{\hat m \hat \nu} \right)^2 \right] G_{\mu\nu}=
\left(\frac{M_g}{\hat m \hat \nu}\right)^2 (\nabla_\mu \nabla_\nu-g_{\mu\nu} \square )  \hat Q^2 +T_{\mu\nu}+T_{\mu\nu}^\varphi+T_{\mu\nu}^{\cal F} \label{geqns}.
\end{multline}
Here ${\cal F}_i$ denotes the partial derivative of ${\cal F}$ with respect to its $i$th argument, $T_{\mu\nu}$ is the energy momentum tensor for the minimally coupled Standard Model fields and 
\begin{eqnarray}
&&T_{\mu\nu}^\varphi = \del_\mu \varphi \del_\nu \varphi-\frac12 g_{\mu\nu} \left( (\del \varphi)^2+m^2 \varphi^2 \right)\\ && T_{\mu\nu}^{\cal F}={\cal F}_3 \del_\mu \varphi \del_\nu \varphi-\mu^4 {\cal F} g_{\mu\nu}
\end{eqnarray}
From \eqref{Qeqn} and \eqref{Qeqns} we obtain the integral constraints
\be
\frac{\mu^4}{m\nu} \langle {\cal F}_1 \rangle=\frac{\int F}{ \textrm{Vol}}, \qquad  \langle R \rangle  \frac{M_g^2 \hat Q}{(\hat m \hat \nu)^2}=\frac{\int \hat F}{ \textrm{Vol}}
\ee
where $\textrm{Vol}=\int \sqrt{-g }d^4 x $ is the spacetime volume and angled brackets denote the spacetime average, $\int X \sqrt{-g }d^4 x=\langle X \rangle \textrm{Vol}$.
 Together these yield the following constraint on the spacetime average of the Ricci scalar
\be \label{globcon}
 \langle R \rangle  \frac{M_g^2 \hat Q}{(\hat m \hat \nu)^2}= \frac{\mu^4}{m\nu} \langle {\cal F}_1 \rangle \frac{\int \hat F}{ \int F}
\ee
This constraint is crucial and nothing more than the sequestering mechanism in action \citeseq. The crucial point is that $\langle R \rangle$ is constrained by the fluxes which correspond to geometric boundary data that can be chosen independently of the UV sector of the theory. Indeed, the flux $\int \hat F$ can be taken to be as small as we like without any violation of naturalness.

Taking traces and spacetime averages of the metric equations of motion \eqref{geqns} we can easily show that
\begin{multline}
M_g^2  \left[1+  \left(\frac{\hat Q}{\hat m \hat \nu} \right)^2 \right] G_{\mu\nu}= T^\text{tot}_{\mu\nu}-\frac14 g_{\mu\nu} \langle T^\text{tot} \rangle -\frac14  M_g^2  \left[1+  \left(\frac{\hat Q}{\hat m \hat \nu} \right)^2 \right] \langle R \rangle g_{\mu\nu}
\end{multline}
where $T^\text{tot}_{\mu\nu}=T_{\mu\nu}+T_{\mu\nu}^\varphi+T_{\mu\nu}^{\cal F} $. After applying the global constraint \eqref{globcon}, we arrive at the following effective gravity equations
\begin{multline} \label{effgrav}
\kappa^2 G_{\mu\nu}=T_{\mu\nu}-\frac14 g_{\mu\nu} \langle T \rangle  +(1+{\cal F}_3)\del_\mu \varphi \del_\nu \varphi -\delta\lambda g_{\mu\nu} \\
-\frac14  \left[ 2{\mu^4}\langle {\cal F}_1 \rangle \kappa^2 \frac{d\hat \sigma}{d\kappa^2}   \frac{\int \hat m \hat \nu \hat F}{ \int  m  \nu F}+ \langle (1+{\cal F}_3) (\del \varphi)^2  \rangle \right] g_{\mu\nu}
\end{multline}
where the effective Planck mass is given by $\kappa^2(\hat \sigma)  =M_g^2  \left[1+\hat \sigma^2\right]$ for constant  $\hat \sigma =\frac{\hat Q}{\hat m \hat \nu}$, and the {\it local} fluctuations\footnote{Since one can easily show that $\langle \delta \lambda \rangle=0$, it is easy to see that $\delta \lambda$ contains no global contribution to the cosmological constant.} in the cosmological constant term are given by
\be \label{dl}
\delta \lambda=\frac12  \left( (\del \varphi)^2+m^2 \varphi^2 \right)+\mu^4 {\cal F}-\left\langle \frac12  \left( (\del \varphi)^2+m^2 \varphi^2 \right)+\mu^4 {\cal F} \right\rangle 
\ee
If we decompose the energy-momentum tensor for matter into its vacuum energy part, $V_{vac}$, and local excitations, $\tau_{\mu\nu}$, as in $T_{\mu\nu}=-V_{vac} g_{\mu\nu}+\tau_{\mu\nu}$ we see that the vacuum energy drops out and we obtain a residual cosmological constant
\be \label{Leff}
\Lambda_\textrm{eff}=\frac14 \langle \tau\rangle +\frac12 \mu^4   \langle {\cal F}_1 \rangle  \kappa^2 \frac{d\hat \sigma}{d\kappa^2} \frac{\int \hat m \hat v \hat F}{ \int m\nu F}+\frac14 \langle (1+{\cal F}_3) (\del \varphi)^2  \rangle
\ee
This quantity is stable against radiative corrections to vacuum energy.  In other words, for a theory cut-off at the scale $M$, 	 although we expect such corrections to go as  $V_{vac} \to V_{vac} +{\cal O}(1) \frac{M^4}{(4\pi)^2} $ \cite{Martin}, we claim that  this will not alter the scale of the residual cosmological constant, $\Lambda_\textrm{eff}\to {\cal O}(1)\Lambda_\textrm{eff}$. To see this let us examine each of the contributions in \eqref{Leff}. The first term, $ \langle \tau \rangle$, is the spacetime average of (the trace of) local matter excitations. By its very definition it receives no corrections from vacuum energy and is small\footnote{In an infinite universe where all localised matter is ultimately diluted away, this will become infinitesimally small.} in a universe that grows large and old, provided matter satisfies the weak energy condition \cite{KP2}. The fluxes $\int m\nu F$,   $\int \hat m \hat v \hat F$ in the second term  are essentially the same as in previous versions of vacuum energy sequester \cite{KPSZ},  rescaled by  $m \nu$ and $\hat m \hat \nu$.  These are purely geometric quantities given entirely by boundary data and not renormalised by radiative corrections to vacuum energy. They can be taken arbitrarily small. 

The contribution from ${\cal F}_1$, or equivalently ${\cal F}$ is also radiatively stable.  To see this, note that $\mu^4{\cal F}$ (or a global subset thereof) plays the role of the cosmological counter term, the bare cosmological constant, whose value is ultimately determined by the geometric global constraint \eqref{globcon}. It therefore scales  with the cut-off as $\frac{M^4}{(4\pi)^2} \sim \mu^4$,  receiving order one radiative  corrections in these units. This means that ${\cal F}$ takes on values of order one, and is corrected to the same order when we include additional loop contributions. Similarly, we have that $\kappa^2 \frac{d\hat \sigma}{d\kappa^2}=\frac{\kappa^2 }{2 \hat \sigma M_g^2} =\frac{\kappa^2/M_{g}^2}{2\sqrt{\kappa^2/M_{g}^2-1}}$. But $\kappa^2$ is the effective gravitational coupling, whose radiative corrections go as $\mu^2$ \cite{Demers}, well below the measured value of $\kappa^2 \sim M_{Pl}^2$. The final term in \eqref{Leff} is also immune to large radiative corrections in essentially the same way as the first term, corresponding to the average of the localised fluctuations in the scalar.

In this set-up, the magnetic duals, $Q$ and $\hat Q$ play a crucial role, akin to the rigid degrees of freedom of the original sequestering proposals \citeseq. The former essentially plays the role of the cosmological counterterm, whilst the latter gives rise to the global geometric constraint that forces the desired cancellation of vacuum energy loops.  Actually, this cannot be the full story because $Q$ is quantised and cannot adjust continuously to compensate for a continuous change in the vacuum energy. However, small changes in the global value of the gauge invariant field $\varphi$ can provide  the extra flexibility required.

We might also be concerned that the mechanism for cancelling  vacuum energy also does away with inflation.  It was already shown that this was not the case in generic sequestering proposals \cite{KP2}, and we see the same here. The key point, of course, is that the value of the  inflationary potential during slow roll represents a local excitation at early times. Indeed, our effective gravity equation \eqref{effgrav} contains an explicit contribution from {\it local} fluctuations in the cosmological constant, $\delta \lambda$. given by \eqref{dl}.  Further, the spacetime average in \eqref{dl} is negligible in a universe that grows old and large. Thus, alongside \eqref{vphieqn}, we see that the dynamics of inflation goes through essentially as in \cite{AKL}.  There will be small corrections of order $\frac{\bar m^2}{m^2}$ coming from the breaking of the gauge symmetry \eqref{symmetry}, and it would be of interest to explore the observational consequences of these as a smoking gun for the sequestering mechanism in Nature.

Finally, let us remark on the corrections coming from inflaton couplings to the Standard Model which we neglected in our analysis.  These do not alter our qualitative results. To see this, note that such corrections endow the matter Lagrangian with dependence on $\phi=\varphi -\frac{Q}{m}$, or equivalently, $ {\cal L}_m \to {\cal L}_m(g^{\mu\nu} ,  \phi, \Phi)$.  This only impacts equations \eqref{Qeqn} and \eqref{vphieqn}, and in each case amounts to trading ${\cal F}_1 \to {\cal F}_1-\frac{\nu}{\mu^4} \frac{\partial {\cal L}_m}{\partial \phi}$. When we take the spacetime average to evaluate the residual cosmological constant, any corrections coming from localised matter excitations contained in $\frac{\partial {\cal L}_m}{\partial \phi}$ will be negligible.

\section{Discussion}
In this paper we have shown how a pair of  field theory monodromies, with deformations motivated by inflation and quantum gravity, naturally incorporates a mechanism for stabilising the observed cosmological constant, protecting it from large radiative corrections to vacuum energy. This cancellation goes through the mechanics of sequestering \citeseq~ and suggests that monodromy constructions within string theory could be adapted to allow for a radiatively stable cosmological constant at low energies.  It is important that the two monodromies operate at hierarchically different scales: for the low scale  inflationary monodromy, the inflaton moves according  to slow roll, while its magnetic dual plays the leading role in the cosmological counterterm required to cancel radiative corrections to vacuum energy. The high scale dilaton monodromy, in contrast, is held rigid. Only the magnetic dual plays any role forcing the desired global constraint on the geometry. The rigidity of the dilaton sector also  avoids issues with experimental tests of General Relativity \cite{test}.

We can think of monodromy as the natural way in which we would extend  low energy sequestering models into the UV, such that the cancellation mechanism described above might have been anticipated.  
To see this explicitly consider for definiteness and simplicity the local formulation  of sequestering introduced in \citep{KPSZ} (although we note that much of what we say here can easily be adapted to the improved model \cite{KP4} designed to sequester vacuum energy contributions from graviton loops). The action introduced in \cite{KPSZ} is given by
\begin{multline} \label{seqaction}
S=\int d^4 x \sqrt{ g} \left[ \frac{\kappa^2(x)}{2}  R   -\Lambda(x)+{\cal L}_m(g^{\mu\nu} , \Phi) \right]+\int \sigma \left(\frac{  \Lambda}{ \mu^4}\right)  F
+ \int \hat \sigma\left(\frac{\kappa^2}{M_g^2} \right)  \hat F \, .
\end{multline}
where $F=\frac{1}{4!}  F_{\mu\nu\alpha\beta}dx^\mu dx^\nu dx^\alpha dx^\beta$ and its hatted counterpart correspond to four-form field strengths. This theory contains a cosmological potential $\Lambda$ and a dilaton $\kappa$ each of whom are held rigid by the dynamics of the three form fields but whose global variation ensures the cancellation of vacuum energy contributions coming from matter loops.  Typically we assume that $M_g \sim M_{Pl}$ and that $\mu$ is around the cut-off. We now reparametrise the theory \eqref{seqaction}, introducing the fields $\phi=\nu \sigma$, $\hat \phi=\hat \nu \hat \sigma$ then defining the potentials $\Lambda=\mu^4 \theta(\sigma)$,  $\kappa^2=M_g^2 \hat \theta (\hat \sigma)$. After  rescaling $F \to m \nu F$, $\hat F \to \hat m \hat \nu \hat F$, we obtain,
\begin{multline} \label{seqaction2}
S=\int d^4 x \sqrt{ -g} \left[ \frac{M_g^2}{2}  \hat \theta \left(\frac{\hat \phi}{\hat \nu} \right) R   -\mu^4 \theta \left(\frac{ \phi}{ \nu} \right)+ {\cal L}_m(g^{\mu\nu} , \Phi) \right]
 +\int m \phi  F+ \int \hat m \hat \phi   \hat F \, .
\end{multline}
The rigidity of the scalars $\phi$ and $\hat \phi$ can be relaxed by adding canonical kinetic terms for the scalars and the four-forms, 
\begin{multline} \label{seqaction3}
S=\int d^4 x \sqrt{ -g} \Bigg[ -\frac{1}{2. 4!} F_{\mu\nu\alpha\beta}^2-\frac12 (\partial \phi)^2
+\frac{m}{4!} \phi \frac{\epsilon^{\mu\nu\alpha\beta}}{\sqrt{-g}} F_{\mu\nu\alpha\beta}
-\frac{1}{2. 4!} \hat F_{\mu\nu\alpha\beta}^2-\frac12 (\partial \hat \phi)^2+\frac{\hat m}{4!} \hat \phi \frac{\epsilon^{\mu\nu\alpha\beta}}{\sqrt{-g}} \hat F_{\mu\nu\alpha\beta}     \\+\frac{M_g^2}{2}  \hat \theta \left(\frac{\hat \phi}{\hat \nu} \right) R   -\mu^4 \theta \left(\frac{ \phi}{ \nu} \right)+ {\cal L}_m(g^{\mu\nu} , \Phi)\Bigg]
\end{multline}
On the first two lines above, we see two copies of the original Kaloper-Sorbo theory, coupled through gravity, with deformations that break the corresponding gauge symmetries on the last line. When this theory is written in  terms of scalars only, after integrating out the four-form field strengths, we see that we have two massive scalars, of mass $m$ and $\hat m$.
The original Lagrangian for vacuum energy sequestering \eqref{seqaction2} is recovered at low energies below these mass scales, where we decouple the local fluctuations in the scalars whilst retaining their rigid deformations. One can explicitly show that  this generalised form of vacuum energy sequestering does indeed sequester the vacuum energy successfully.  This is because the vacuum energy source, being at infinite wavelength, only sees the low energy effective theory below the two mass scales, which is, of course, the original theory proposed in \citep{KPSZ}. 

Although we have assumed that the two axion sectors and the matter sector only mix gravitationally, it is natural to ask whether or not we can relax this, and if so, to what extent?  A detailed answer to these questions requires deeper investigation but for now let us make a brief comment. There are essentially two things  that we must consider: the observational consequences of any additional mixings and the implications for the cancellation of radiative corrections to vacuum energy. For  example, from an observational standpoint, we might be concerned if the light axion were to couple non-minimally to gravity, as does the heavy axion.  The light axion is sufficiently heavy for this to leave local gravity tests unaffected although there might be some implications for cosmology along the lines suggested in \cite{stein}. Regarding the robustness of the cancellation mechanism, it is important to retain the global geometrical constraint without any contamination from UV sensitive sectors, as explained in \cite{etude}.  This suggests that problems could arise if the heavy axion mixes directly with matter fields. 
 
\begin{acknowledgments}
\vskip.5cm

{\bf Acknowledgments}: 
  A.P. would like to thank N. Kaloper, B. El Manoufi, S. Nagy, F. Niedermann, P. Saffin and D. Stefanyszyn for useful discussions.  Note that some of the technicalia on sequestering monodromies were identified in discussions with N. Kaloper. A.P is  funded by a Leverhulme Trust
  Research Project Grant and an STFC Consolidated Grant.
\end{acknowledgments}

\end{document}